\documentclass[a4paper,12pt]{article}

\usepackage{latexsym}
\usepackage{amsfonts} 
\usepackage[mathscr]{euscript}
\usepackage{amsmath,mathrsfs,bm,amssymb,color}

\title{\textsf{Scale Dependence of the Retarded van  der Waals Potential }}
\date{\empty}

\author{
Tadahiro Miyao\footnotemark[1]
  and  Herbert Spohn\footnotemark[2]\\ 
\footnotemark[1] {\it Department of Mathematics,}
{\it Hokkaido University,}\\
{\it Sapporo 060-0810, Japan}\\
\footnotemark[2] {\it Zentrum Mathematik,}
{\it Technische Universit\"at M\"unchen,}\\ 
{\it  D-85747 Garching, Germany}\\
$$
\footnotemark[1] {\tt miyao@math.sci.hokudai.ac.jp,}
\footnotemark[2] {\tt spohn@ma.tum.de} }

\newcommand{\one}{{\mathchoice {\rm 1\mskip-4mu l} {\rm 1\mskip-4mu l}
{\rm 1\mskip-4.5mu l} {\rm 1\mskip-5mu l}}}

\newcommand{\ex}{\mathrm{e}}
\newcommand{\D}{\mathrm{dom}}

\newcommand{\im}{\mathrm{i}}
\newcommand{\Fock}{\mathfrak{F}}

\newcommand{\la}{\langle}
\newcommand{\ra}{\rangle}
\newcommand{\Tr}{\mathrm{Tr}}

\newcommand{\BbbR}{\mathbb{R}}

\newcommand{\BbbC}{\mathbb{C}}
\newcommand{\vepsilon}{\varepsilon}
\newcommand{\vphi}{\varphi}

\newcommand{\Hf}{H_{\mathrm{f}}}

\newcommand{\dm}{\mathrm{d}}

\newcommand{\no}{\nonumber \\}

\begin{document}

\newtheorem{define}{Definition}[section]
\newtheorem{Thm}[define]{Theorem}
\newtheorem{Prop}[define]{Proposition}
\newtheorem{lemm}[define]{Lemma}
\newtheorem{rem}[define]{Remark}
\newtheorem{assum}{Condition}
\newtheorem{example}{Example}
\newtheorem{coro}[define]{Corollary}
\newtheorem{conj}[define]{Conjecture}

\maketitle
\noindent
\textit{We dedicate our contribution to Elliott Lieb with greatest admiration and deep gratitude for what he has
taught us, and whole generations, about quantum mechanics and statistical physics.} \\ 
{\abstract{\noindent We study the ground state energy for a system of two hydrogen atoms coupled to the quantized Maxwell field in the limit $\alpha \to 0$ together with the relative distance between the atoms increasing as $\alpha^{-\gamma} R$, $\gamma > 0$. In particular we determine explicitly the crossover function from the $R^{-6}$ van der Waals potential to the $R^{-7}$ retarded van der Waals potential, which takes place at scale $\alpha^{-2} R$.}}

%%%%%%%%%%%%%%%%%%%%%%%%%%%%%%%%%%%%%%%%%%%%%%%%%%%%%%%%

\section{Introduction}
In a now very  famous contribution,  Casimir and   Polder \cite{CP} investigate the ground
state energy, $E(R)$, of a system of two hydrogen atoms for which the two 
immobile nuclei are separated by  a distance $R$ and the two spinless
electrons are coupled through the quantized Maxwell field
according to non-relativistic QED. In the approximation where the quantum
fluctuations of the Maxwell field are ignored, only the electrostatic
Coulomb interaction remains. In this  case $E(R)-E(\infty)\approx
-R^{-6}$, the ubiquitous van der Waals potential, which has been  discovered
on  thermodynamic grounds way before the advent of quantum mechanics.
The $R^{-6}$ behavior is well understood quantum mechanically \cite{MS}
and has been proved in great generality by Lieb and Thirring \cite{LT}.
Casimir and Polder use fourth order perturbation theory to argue that
because of retardation  effects the true asymptotic behavior  is in fact
$E(R)-E(\infty)\approx -R^{-7}$ for large $R$.
Their argument has been reworked many times and extended to arbitrary
atoms and molecules, see for example  \cite{MK,Ex,Milonni,FS0,FS,Boyer}. It is generally agreed
that for two \textit{neutral} molecules $A, B$ it holds 
\begin{align}
E(R)-E(\infty)\cong -\frac{23}{4\pi} \alpha_A \alpha_B R^{-7} \label{CP1}
\end{align} 
for large $R$. Here $\alpha_A$, $\alpha_B$, are the electric dipole
moments of molecule $ A$, $B$.
The numerical prefactor is universal ($23/4\pi$ is the value in Gaussian  units).
 
(\ref{CP1}) is based on perturbation theory and thus holds only for small
coupling.
With improved experimental techniques, there has been a renewed interest
to explore a wider regime. One still finds the $R^{-7}$ power law, but 
the prefactor is now a bilinear form in the electric and magnetic dipole
moments. To be consistent, in principle,  these moments have to be computed  for the
single molecule in isolation but still coupled to its own quantized
radiation field. All atomic/molecular properties appear through  the
electric and magnetic dipole moments.
 As in (\ref{CP1}), the remaining coefficients are universal. In
 particular the coefficient $23/4\pi$ for electric dipole-electric
 dipole contribution  persists.
To mention only the most recent work: in \cite{BK,SW} the retarded van der
Waals potential
is computed in the framework of macroscopic QED. The approach in \cite{MS1}
is based on the standard non-relativistic QED hamiltonian, but uses the representation in terms of a
functional integral. Conceptually this has the advantage that
$E(\infty)$ is subtracted without error and that $1/R$ turns into
a small parameter explicitly showing up in the action. Thus $1/R$
can be used as an expansion parameter, which is more physical than the conventional
 coupling strength to the Maxwell field. 

In the framework of non-relativistic QED the existence of a ground state,
for arbitrary $R$ and coupling strength, has been established in the
break through contribution of Griesemer, Lieb, and Loss \cite{GLL}. 
To determine the leading, large $R$ asymptotics of $E(R)$ seems to be a
difficult problem, even for small, but fixed, coupling. In view of this
situation we develop here a novel approach somewhat closer in spirit to the
original  Casimir-Polder considerations.
As  interaction strength we use the fine structure constant
$\alpha$
and regard the ground state energy, $E(R)=E_{\alpha}(R)$, as depending both on $R$ and
$\alpha$. We then study the limit of small coupling with an
approximately adjusted scale of $R$, more precisely we consider the
limit
\begin{align}
E_{\alpha}(\alpha^{-\gamma} R)-E_{\alpha}(\infty)\,,\quad \gamma \geq 0, \label{CP2}
\end{align} 
as $\alpha\to 0$. Depending on the value $\gamma$ distinct features of
$E_{\alpha}(R)$ will become visible. In particular, we will find an explicit formula 
for the crossover from $R^{-6}$ to $R^{-7}$, which occurs at scale $\alpha^{-2}$.

In Section 2 we define the hamiltonians and provide an overview on the dependence on $\gamma$.
There are two special values. At $\gamma = 1$ one crosses from core dominated behavior  to the $- R^{-6}$ van der Waals and at 
$\gamma = 2$ one crosses from $- R^{-6}$ to $- R^{-7}$. The corresponding crossover function is computed 
explicitly and seems to be novel. Sections 3 and 4 provide proofs and point out open problems.\bigskip\\  
\textbf{Acknowledgements}: We thank Stefan Teufel for raising the question on scale dependence,
Mario Koppen for discussions on the asymptotic analysis, and Ales\-sandro Pizzo for pointing at Nelson's theorem on resolvent convergence.

%%%%%%%%%%%%%%%%%%%%%%%%%%%%%%%%%%%%%%%%%%%%%%%%%%%%%%%%%%%%

\section{Hamiltonians  and main results}\setcounter{equation}{0}

Let us first  consider a single hydrogen atom with an infinitely heavy nucleus located at the origin. The nucleus has charge $e$, $e>0$,
the electron has charge $-e$. We will use units in which $\hbar=1, c=1$, and the bare mass of the electron $m=1$. 
Then the fine-structure constant is $\alpha = e^2/4\pi$.
 Let $x,p$ be position and momentum of the spinless electron. Then the non-relativistic QED hamiltonian for this system reads 
\begin{align}
H_{1,\alpha}=\tfrac{1}{2}:\hspace*{-3pt} \big(p-eA(x)\big)^2\hspace*{-3pt}: - e^2V_{\vphi}(x)+\Hf\,.\label{1PL}
\end{align}
The electrons and the nuclei are assumed to have the same  prescribed  charge distribution $\vphi$ with the following properties: $\vphi$ is  normalized,
$\int \mathrm{d}x\, \vphi(x)=1$, rotation invariant,
$ \vphi(x)=\vphi_{\mathrm{rad}}(|x|)$,  and of rapid decrease. Denoting
Fourier transform by  $\hat{\vphi}$,  the potential 
$V_{\vphi}$ is the smeared Coulomb potential
\begin{align}
V_{\vphi}(x)=\int_{\BbbR^3}\mathrm{d}k\, |\hat{\vphi}(k)|^2|k|^{-2} \ex^{-\im k\cdot x}\,.
\end{align}
$A(x)$ is the quantized  vector potential and $\Hf$  is the field energy. These are defined through a two-component Bose field 
$a(k, \lambda), k\in \BbbR^3, \lambda=1,2,$ with commutation relation
\begin{align}
[a(k, \lambda), a(k', \lambda')^*]= \delta_{\lambda\lambda'}\delta(k-k')\,.
\end{align}
Explicitly
\begin{align}
\Hf=\sum_{\lambda=1,2}\int_{\BbbR^3}\mathrm{d}k\, \omega(k)a(k, \lambda)^*a(k, \lambda)
\end{align}
with dispersion  relation
\begin{align}
\omega(k)=|k|
\end{align}
and 
\begin{eqnarray}
&&\hspace{-10pt}A(x)=\sum_{\lambda=1,2}\int_{\BbbR^3}\mathrm{d}k\,\hat{\vphi}(k) \frac{1}{\sqrt{2\omega(k)}}\vepsilon(k, \lambda)\big(\ex^{\im k\cdot x}a(k, \lambda)+\ex^{-\im k\cdot x}a(k, \lambda)^*\big)\nonumber\\
&&\hspace{18pt}=A^+(x)+A^-(x)
\end{eqnarray}
with the standard dreibein $\vepsilon(k,1), \vepsilon(k, 2),
\hat{k}=k/|k|$. 
$:\cdot:$ denotes normal ordering, which will be of use later on.
Thus the Hilbert space for $H$ is 
\begin{align}
\mathcal{H}=L^2(\BbbR^3_x)\otimes \mathfrak{F}\,,
\end{align} 
where $\mathfrak{F}$ is the bosonic Fock space over $
L^2(\BbbR^3)\otimes \BbbC^2$.
From the quantization  of the classical system of charges coupled to
the Maxwell field it follows that for the smearing of $A(x)$ and of
$V_{\vphi}$ the same charge distribution has to be used. We refer to
\cite{Spohn} for details.
The ground state energy of $H_{1, \alpha}$ is denoted by $E_{1, \alpha}$.

To investigate  the van der Waals potential one considers  two hydrogen atoms, one
located at $0$ and the other at $r=(0,0,R), R\ge 0$. It will be  convenient to
define the position of the second electron relative to $r$. Then $x_1,
x_2+r$ are positions  and $p_1, p_2$ the momenta of the two
electrons. The two-electron hamiltonian reads
\begin{align}
H_R=&\tfrac{1}{2}:\hspace*{-3pt}\big(p_1-eA(x_1)\big)^2\hspace*{-3pt}:-e^2V_{\vphi}(x_1)+\tfrac{1}{2}:\hspace*{-3pt}\big(p_2-eA(x_2+r)\big)^2\hspace*{-3pt}:-e^2V_{\vphi}(x_2)\nonumber \\[1ex]
&+\Hf+e^2 V_R(x_1, x_2)\label{FullHami}
\end{align}
with the interaction potential
\begin{align}
V_R(x_1, x_2)&= -V_{\vphi}(x_1-r)-V_{\vphi}(x_2+r)+V_{\vphi}(r)+V_{\vphi}(r+x_2-x_1)\nonumber \\
&=\int_{\BbbR^3}\mathrm{d}k\, |\hat{\vphi}(k)|^2\ex^{\im k\cdot r}
|k|^{-2}
(1-\ex^{-\im k\cdot x_1})(1-\ex^{\im k\cdot x_2})\,.
\end{align}
$H_R$  acts on the Hilbert space $L^2(\BbbR^3_{x_1})\otimes
L^2(\BbbR^3_{x_2})\otimes \mathfrak{F}$. $H_R$ has a unique ground
state with energy $E_{\alpha}(R)$. It is known that $\lim_{R\to \infty}
E_{\alpha}(R)=2E_{1,\alpha}$. 

We plan to study $E_{\alpha}(\alpha^{-\gamma} R)$ in the limit of small $\alpha$ and consider 
first the hydrogen atom.
In the limit $\alpha\to 0$ the Bohr radius is order $\alpha^{-1}$ and
the energy is order $-\alpha^{2}$. Hence it is convenient to switch  to
atomic coordinates which amounts to the unitary transformation 
\begin{align}
&U^* a(k, \lambda)U=\alpha^{-3} a(\alpha^{-2}k, \lambda)\,,\ \ 
U^* x U=\alpha^{-1} x\,,\ \  U^* p U =\alpha p\,, \nonumber\\[1ex]
& U^* x_j U=\alpha^{-1}
 x_j\,,\ \ U^* p_j U=\alpha p_j\,, \ \ j=1,2\,.
\end{align}  
Then
\begin{align}
U^* H_{1,\alpha} U=\alpha^2 \Big(
\tfrac{1}{2}:\hspace*{-3pt}\big(
p-\sqrt{4\pi} \alpha^{3/2} A_{\alpha}(x)
\big)^2\hspace*{-3pt}:
-V_{\alpha}(x)+\Hf
\Big)
\end{align} 
with 
\begin{align}
V_{\alpha}(x)=4\pi \int_{\BbbR^3}\dm k\,  |\hat{\vphi}(\alpha  k)|^2 |k|^{-2}\ex^{-\im
 k\cdot x }
\end{align}
and 
\begin{align}
A_{\alpha}(x)&=\sum_{\lambda=1,2} \int_{\BbbR^3}\dm k\, 
\hat{\vphi}(\alpha^2 k) \frac{1}{\sqrt{2|k|}}\vepsilon(k, \lambda) 
\big(
\ex^{\im \alpha k\cdot x} a(k, \lambda)+\ex^{-\im \alpha k\cdot x} a(k, \lambda)^*
\big)\no
&=A_{\alpha}^+(x)+A_{\alpha}^-(x)\,.
\end{align} 
We note that
\begin{align}
\alpha^3[A_{\alpha}^+(x), A_{\alpha}^-(x)]=\alpha^3 \int_{\BbbR^3}\dm
 k\, |\hat{\vphi}(\alpha^2 k)|^2 |k|^{-1}=\mathcal{O}(\alpha^{-1})\,.
\end{align} 
Thus normal ordering is introduced to subtract these more singular contributions.

Correspondingly the atomic scale hamiltonian
for two hydrogen atoms separated by a distance
$\alpha^{-1} R$ reads
\begin{align}
U^* H_{\alpha^{-1}R} U=&\alpha^2\Big(
\tfrac{1}{2}:\hspace*{-3pt}\big(
p_1-\sqrt{4\pi}\alpha^{3/2} A_{\alpha}(x_1)
\big)^2\hspace*{-3pt}:
+
\tfrac{1}{2}:\hspace*{-3pt}
\big(
p_2-\sqrt{4\pi}\alpha^{3/2} A_{\alpha}(x_2+r)
\big)^2\hspace*{-3pt}:
\no
&-V_{\alpha}(x_1)-V_{\alpha}(x_2)+V_{\alpha, R}(x_1, x_2)+\Hf
\Big)\,. 
\end{align} 
\\ 
%%%%%%%%%%%%%%%%%%%%%%%%%%%%%%%%%%%%%%%%%%%%%%%%%%%%%%%%
\textbf{2.1 The scale $0 \le \gamma \le 1$}\bigskip\\
For $\gamma = 1$, instead of considering merely the ground state energy, 
a more complete picture would be the strong convergence of resolvents. For the smeared Coulomb potentials it holds
\begin{eqnarray}
&&\hspace{-40pt}\lim_{\alpha\to 0} \sup_{x}|V_{\alpha}(x) - |x|^{-1}| = 0\,,\label{1p}\\
&&\hspace{-40pt}\lim_{\alpha\to 0} \sup_{x_1,x_2}\big|V_{\alpha, R}(x_1, x_2) \nonumber\\
&&-\big( -|x_1+r|^{-1}
-|x_2+r|^{-1}+
R^{-1} + |r+x_2-x_1|^{-1}\big)\big| = 0\,.\label{2p}
\end{eqnarray} 
Thus the issue of strong resolvent convergence is reduced to the study of the free particle hamiltonian
\begin{align}\label{free}
T_{1,\alpha} = 
\tfrac{1}{2}:\hspace*{-3pt}\big(
p-\sqrt{4\pi} \alpha^{3/2} A_{\alpha}(x)
\big)^2\hspace*{-3pt}:
+\Hf
\end{align}
and correspondingly for two free electrons, with hamiltonian denoted by $T_{2,\alpha}$. Note that the norm of the coupling function in (\ref{free}) diverges as $\alpha^{-1/2}$. Thus the limit $\alpha \to 0$ is singular. On the other hand
the  recent estimate \cite{BV} of the ground state energy $E^0_{1,\alpha}$ of  $T_{1,\alpha}$ establishes that 
$E^0_{1,\alpha} = -  a_0 + a_3 \alpha+ \mathcal{O}(\alpha^2)$.
For us only the coefficient  $a_0$ 
is of interest, which is given by
\begin{equation}
a_0 =  (2 \pi)^2\big\langle A^+_1(0)\cdot A^+_1(0)\Omega,\big(\tfrac{1}{2}P_\mathrm{f}^2 + H_\mathrm{f}\big)^{-1} A^+_1(0)\cdot A^+_1(0)\Omega \big\rangle
\,.
\end{equation}
Here $\Omega$ denotes the Fock vacuum and $P_\mathrm{f}$ is the field momentum,
\begin{equation}
P_\mathrm{f} =\sum_{\lambda=1,2}\int_{\BbbR^3}\mathrm{d}k\, ka(k, \lambda)^*a(k, \lambda)\,.
\end{equation}
With this information one arrives at 
\begin{conj}\label{ConSR}
In the sense of strong convergence of resolvents,
\begin{eqnarray}
&&\hspace{-16pt}\lim_{\alpha\to 0} T_{1,\alpha}=\tfrac{1}{2}p^2 + H_\mathrm{f} -a_0\,, \\
&&\hspace{-16pt}\lim_{\alpha\to 0} T_{2,\alpha}=\tfrac{1}{2}p_1^2 + \tfrac{1}{2}p_2^2+ H_\mathrm{f} - 2a_0\,.
\end{eqnarray} 
\end{conj}

To our surprise, this limit has apparently never been investigated. In Section 4 we provide some arguments towards the validity of Conjecture \ref{ConSR}.  If it holds, then by (\ref{1p}) and (\ref{2p}) we conclude that
\begin{eqnarray}
&&\hspace{-50pt}\lim_{\alpha\to 0} \alpha^{-2} U^* H_{1,\alpha} U=\frac{1}{2}p^2-\frac{1}{|x|}+\Hf -a_0\no 
&&\hspace{26pt}=H_{\mathrm{hy}}+\Hf - a_0\,,
\end{eqnarray} 
and
\begin{eqnarray}
&&\hspace{-30pt}\lim_{\alpha\to 0} \alpha^{-2} U^* H_{\alpha^{-1}R}U
=\frac{1}{2}p_1^2+\frac{1}{2}p_2^2-|x_1|^{-1}-|x_2|^{-1}-|x_1-r|^{-1}
-|x_2+r|^{-1}\no
&&\hspace{80pt}+R^{-1}+|r+x_2-x_1|^{-1}+\Hf - 2a_0\no
&&\hspace{22pt}= H_{2,R}+\Hf - 2a_0
\end{eqnarray} 
in the sense of strong resolvent convergence.
Denoting the ground state energy of $H_{2,R}$ by $E_{2,R}$, in
particular it holds
\begin{flalign}
&\gamma=1: \hspace*{80pt}\lim_{\alpha\to 0}
 \alpha^{-2}E_{\alpha}(\alpha^{-1}R)=E_{2,R} -2a_0\,.&
\end{flalign} 
Note that 
\begin{align}
\lim_{R\to 0}(E_{2,R}-R^{-1})=E_{\mathrm{he}}
\end{align} 
with $E_{\mathrm{he}}$ the ground state of the helium atom, while
\begin{align}
\lim_{R\to \infty} R^{6}(E_{2,R}-2E_{\mathrm{hy}})=-a_{\mathrm{VW}}\,,
\end{align} 
where $a_{\mathrm{VW}}$ is the strength of the van der Waals potential,
\begin{align}
a_{\mathrm{VW}}=6\int_0^{\infty}\dm t\,  \big|
\tfrac{1}{3} \big\la \psi_0, x\cdot \ex^{-t(H_{\mathrm{hy}}-E_{\mathrm{hy}})}x\psi_0\big\ra
\big|^2\,,
\end{align} 
with $H_{\mathrm{hy}}\psi_0=E_{\mathrm{hy}}\psi_0, \ E_{\mathrm{hy}}=-1/2$.
Thus we conclude that on the distance scale $\alpha^{-1}R$ the energy
$\alpha^2 E_{2,R}$ describes the crossover to the $-R^{-6}$ potential.

For completeness we list the even smaller distance scales,
\begin{flalign}
&\gamma=0: \hspace*{62pt} E_{\alpha}(R) \cong \alpha V_{\alpha}(R)+\alpha^2
 (E_{\mathrm{he}} -2a_0)\,,&\\[1ex]
&0<\gamma<1: \hspace*{40pt}E_{\alpha}(\alpha^{-\gamma}R)\cong
 \alpha^{1+\gamma}R^{-1}+\alpha^2(E_{\mathrm{he}} -2a_0)\,.&
\end{flalign}
\\ 
%%%%%%%%%%%%%%%%%%%%%%%%%%%%%%%%%%%%%%%%%%%%%%%%%%%%%%%%
\textbf{2.2 The scale} $\gamma \ge 1$\bigskip\\
To go beyond the distance scale $\alpha^{-1}R$ is a more difficult
problem and we have only  partial results. We expect that the range
$1<\gamma <2$ is dominated by the van der Waals potential, {\it i.e.}
\begin{flalign}\label{gamma1}
&1<\gamma<2: \hspace*{40pt}E_{\alpha}(\alpha^{-\gamma}R)\cong \alpha^{6\gamma-4} a_{\mathrm{VW}}
 R^{-6}+2E_{1, \alpha}\,.&
\end{flalign}

The retardation of the van der Waals potential first appears at scale
$\alpha^{-2}$. More precisely
\begin{flalign}\label{gamma2}
&\gamma=2:\hspace*{62pt} E_{\alpha}(\alpha^{-2}R)\cong -\alpha^8 h_{\mathrm{co}}(R)+2E_{ 1,\alpha}\,,&
\end{flalign} 
where the crossover function $h_{\mathrm{co}}$ is defined by 
\begin{eqnarray}
&&\hspace{-45pt}h_{\mathrm{co}}(R)\no
&&\hspace{-40pt}= \pi^{-1}\int_{0}^{\infty}\dm u 
\big(
\tfrac{1}{3}\big\la \psi_0, x\cdot (H_{\mathrm{hy}}-E_{\mathrm{hy}})\big((H_{\mathrm{hy}}-E_{\mathrm{hy}})^2+(u/2)^2\big)^{-1}x\psi_0\big\ra
\big)^2\ex^{- R u }\no
&&\hspace{0pt}\times \Big\{
2^{-3}R^{-2}u^4+2^{-1}R^{-3}
 u^3+5\cdot 2^{-1}R^{-4}
 u^2 +6R^{-5} u +6R^{-6}
\Big\}\,. \label{cofunction}
\end{eqnarray} 
At small distances
\begin{align}
h_{\mathrm{co}}(R)\cong a_{\mathrm{VW}} R^{-6}, \ \ \mbox{as $R\to 0$}\,,
\end{align} 
and at large distances
\begin{align}
h_{\mathrm{co}}(R)\cong a_{\mathrm{CP}} R^{-7}, \ \ \mbox{as $R\to \infty$}\,.
\end{align} 
Here the strength of the retarded van der Waals potential is 
\begin{align}
a_{\mathrm{CP}}= \frac{23}{4\pi}(\alpha_{\mathrm{hy}})^2\,,\ \ \
 \alpha_{\mathrm{hy}} =\frac{2}{3}\big\la
\psi_{\mathrm{hy}}, x\cdot (H_{\mathrm{hy}}-E_{\mathrm{hy}})^{-1}x\psi_{\mathrm{hy}}
\big\ra=\frac{9}{2}\,.
\end{align} 
We conclude that at scale $\alpha^{-2}R$ the ground state energy crosses
from  the van der Waals potential to the retarded one as specified  by
$h_{\mathrm{co}}$.

At even larger scales one expects the exact power low $R^{-7}$,
\begin{flalign}\label{gamma3}
&\gamma>2:\hspace*{60pt}
 E_{\alpha}(\alpha^{-\gamma}R)\cong -\alpha^{7\gamma-6}a_{\mathrm{CP}}
 R^{-7}+2E_{ 1,\alpha}\,.&
\end{flalign} 

The hydrogen atom ground state has been estimated up to
$\mathcal{O}(\alpha^5 \log \alpha^{-1})$ \cite{BCVV} based on  a method
originally devised by Hainzl and Seiringer \cite{HS}. It is rather natural
 to use similar methods for the case of two hydrogen atoms. If more modestly
 we strive  for  a precision of order $\alpha^3$, then the scale will be
 limited to $\alpha^{-6/5}$, unless there is a more direct way to
 accomplish the subtraction.
At scale $\alpha^{-2}$ the first term is order $\alpha^8$. There is no
hope
to control $E_{1, \alpha}$ with  such a precision and one has to look
for alternative schemes.

%%%%%%%%%%%%%%%%%%%%%%%%%%%%%%%%%%%%%%%%%%%%%%%%%%%

\section{The $\gamma=2$ crossover function}\setcounter{equation}{0}

Our main goal is to derive the crossover function of
(\ref{cofunction}). The starting point is the functional integral representation
of  $E_{\alpha}(R)-E_{\alpha}(\infty)$, see \cite{MS1}. In this paper we
consider only the second cumulant of the action, assuming that higher
cumulants decay at least as $R^{-8}$ for $R\to \infty$.
The functional integration is defined with respect to two independent
ground state processes for the hamiltonian $H$ of (\ref{1PL}). On the basis of our
conjecture, for small
coupling we replace $H$ by 
\begin{align}
\alpha^2\big(H_{\alpha}+\Hf -a_0\big)\,,\quad  H_{\alpha} =\tfrac{1}{2}p^2-  V_{\alpha}(x)\,.
\end{align} 
Denoting both approximations by $[\cdots]_{\mathrm{cu}}$ we have 
\begin{align}
[E_{\alpha}(R)-E_{\alpha}(\infty)]_{\mathrm{cu}}= - 2(2 \pi\alpha)^2\big(
I_2(R)+I_3(R)+I_4(R)
\big)\,,\label{CumuRep}
\end{align}  
where the coefficients are given in (39) resp. (47), (52) of \cite{MS1} with the
understanding that $H$ is replaced by $\alpha^2\big(H_{\alpha}+\Hf -a_0\big)$. (\ref{CumuRep})
should be regarded as the definition of the left hand side.
Required is an asymptotic analysis of the integrals $I_2$, $I_3$, and
$I_4$.
In fact, in the scalings of (\ref{gamma1}), (\ref{gamma2}),
(\ref{gamma3}), the contribution of $I_2$ and $I_3$ vanish as $\alpha\to 0$.
The details are lengthy and will  not be recorded here. In the following we focus only on the relevant
contribution $I_4$.

For notational simplicity, we replace $H_{\alpha}-\inf \mathrm{spec}(H_{\alpha})$ by
$H_{\alpha}$  in the remainder of this section.
Then the ground state $\psi_\alpha$ is defined through $H_{\alpha}\psi_\alpha = 0$.
Similarly  $H_{\mathrm{hy}}-E_{\mathrm{hy}}$ is replaced by $H_{\mathrm{hy}}$.
 Hence $H_{\mathrm{hy}}\psi_{\mathrm{hy}}=0$. According to (52) of \cite{MS1},
$I_4(R)$ is defined by
\begin{eqnarray}
&&\hspace{-40pt}I_4(R)=\alpha^6 \int_{\BbbR^3}\dm t_1\dm t_2\dm t_3 \sum_{\lambda_1, \lambda_2}
\int_{\BbbR^6}\dm k_1\dm k_2 |\hat{\varphi}(\alpha^2 k_1)|^2
 |\hat{\varphi}(\alpha^2 k_2)|^2 \ex^{\im
 (k_1+k_2)\cdot r \alpha^{2}}\no
&&\hspace{10pt}\times  \tfrac{1}{4}\omega_1\omega_2\,  \ex^{-\omega_1|t_1+t_2+t_3|}
 \ex^{-\omega_2|t_3|}\no[1ex]
&&\hspace{10pt}\times\big\la
\psi_{\alpha}, (\vepsilon_1\cdot x) H_{\alpha}\ex^{-\im k_1\cdot x \alpha}
 H_{\alpha}^{-2}\, 
\ex^{-|t_1| H_{\alpha}}
\, \ex^{-\im k_2\cdot x \alpha }H_{\alpha} (\vepsilon_2\cdot x)
\psi_{\alpha}
\big\ra\no[1ex]
&&\hspace{10pt}\times\big\la
\psi_{\alpha}, (\vepsilon_1\cdot x) H_{\alpha} \ex^{\im k_1\cdot x \alpha }
 H_{\alpha}^{-2}\, 
\ex^{-|t_2| H_{\alpha}}
\, \ex^{\im k_2\cdot x \alpha}H_{\alpha}  (\vepsilon_2\cdot x)
\psi_{\alpha}
\big\ra\,,
\end{eqnarray}  
where now the inner product is in $L^2(\BbbR^3_x)$.

\begin{Prop}\label{AssertionI4}
Assume that the smearing function $\vphi$ is radial, continuous, and of compact support.\medskip\\
(i) Let $1 < \gamma < 2$. Then
\begin{align}
\lim_{\alpha\to 0}\alpha^{6-6\gamma} (4\pi)^{-2}
 I_4(\alpha^{-\gamma}R)
=a_{\mathrm{VW}} R^{-6}\,.\medskip
\end{align} 
(ii) Let $\gamma = 2$. Then 
\begin{equation}
\lim_{\alpha\to 0}\alpha^{-6}I_4(\alpha^{-2}R) = h_{\mathrm{co}}(R)\,.\medskip
\end{equation}
(iii) Let $\gamma > 2$. Then
\begin{equation}
\lim_{\alpha\to 0} \alpha^{8 -7\gamma }(4\pi)^{-2}
 I_4(\alpha^{-\gamma}R)
=\frac{23}{4\pi}(a_{\mathrm{hy}})^2 R^{-7}\,.
\end{equation} 
\end{Prop}
Using (\ref{CumuRep}) the proposition supports the claims of Section 2.\bigskip\\
\textbf{Proof}: 
For better readability we subdivide our proof into several steps. 
But before  we remark that, within the current proof, compact support of $\varphi$ is required.
\medskip\\
\textbf{Step 1} (Rewriting). We  scale  $k_j\leadsto \alpha^{\gamma-2}k_j$ and $t_3\leadsto \alpha^{2-\gamma}t_3$.  Then
\begin{eqnarray}
&&\hspace{-40pt}I_4(\alpha^{-\gamma}R)\no
&&\hspace{-20pt}=\alpha^{7\gamma-8}\int_{\BbbR^3}\dm t_1\dm t_2\dm t_3 \sum_{\lambda_1, \lambda_2}
\int_{\BbbR^6}\dm k_1\dm k_2  |\hat{\varphi}(\alpha^{\gamma}k_1)|^2
 |\hat{\varphi}(\alpha^{\gamma} k_2)|^2\no
 &&\hspace{20pt}\times\ex^{\im
 (k_1+k_2)\cdot r}
\tfrac{1}{4}\omega_1\omega_2\,  \ex^{-\alpha^{\gamma-2}\omega_1|t_1+t_2+\alpha^{2-\gamma}t_3|}
 \ex^{-\omega_2|t_3|}\no
 &&\hspace{20pt}\times\big\la
\psi_{\alpha}, (\vepsilon_1\cdot x) H_{\alpha} \ex^{-\im \alpha^{\gamma-1} k_1\cdot x}
 H_{\alpha}^{-2}\, 
\ex^{-|t_1| H_{\alpha}}
\, \ex^{-\im  \alpha^{\gamma-1}k_2\cdot x}H_{\alpha} (\vepsilon_2\cdot x)
\psi_{\alpha}
\big\ra
\no[1ex]
&&\hspace{20pt}\times\big\la
\psi_{\alpha}, (\vepsilon_1\cdot x) H_{\alpha} \ex^{\im \alpha^{\gamma-1}k_1\cdot x}
 H_{\alpha}^{-2}\, 
\ex^{-|t_2| H_{\alpha}}
\, \ex^{\im \alpha^{\gamma-1}k_2\cdot x}H_{\alpha} (\vepsilon_2\cdot x)
\psi_{\alpha}
\big\ra\,.
\label{I4}
\end{eqnarray} 
Let us  note the following equality,
\begin{align}
&\int_{\BbbR^3}\dm t_1\dm t_2 \dm  t_3\, \ex^{-\omega_1
 \alpha^{\gamma-2}|t_1+t_2+\alpha^{2-\gamma }t_3|} \ex^{-\omega_2|t_3|}
\ex^{-\lambda_1|t_1|}\ex^{-\lambda_2|t_2|}\no
&=(2\pi)^{-1} \alpha^{2-\gamma}\int_{\BbbR}\dm u
 \frac{2\omega_1}{\omega_1^2+\alpha^{4-2\gamma}u^2}\cdot
\frac{2\omega_2}{\omega_2^2+\alpha^{4-2\gamma}u^2}\cdot
\frac{2\lambda_1}{\lambda_1^2+u^2}\cdot
\frac{2\lambda_2}{\lambda_2^2+u^2}\,,
\end{align} 
which is proven by using the Fourier transform
\begin{align}
(2\pi)^{-1}\int_{\BbbR} \dm u \, \ex^{-\im
 ut}\frac{2\omega}{\omega^2+u^2}=\ex^{-\omega|t|}\,.
\end{align}
Viewing $\lambda_1$ and $\lambda_2$ as spectral parameters for $H_\alpha$, one arrives at 
\begin{eqnarray}
&&\hspace{-35pt}I_4(\alpha^{-\gamma}R)
= \alpha^{6\gamma-6}(2\pi)^{-1}\int_{\BbbR}\dm u \int_{\BbbR^6}\dm k_1\dm k_2\, 
|\hat{\varphi}(\alpha^{\gamma}k_1)|^2 |\hat{\varphi}(\alpha^{\gamma} k_2)|^2 \no
&&\hspace{-10pt}\times\ex^{\im
 (k_1+k_2)\cdot r}\frac{k_1^2}{k_1^2+\alpha^{4-2\gamma}u^2}\cdot
\frac{k_2^2}{k_2^2+\alpha^{4-2\gamma}u^2}\no
&&\hspace{-10pt}\times2\big\la
\psi_{\alpha}, (\vepsilon_1\cdot x) H_{\alpha}\ex^{-\im k_1\cdot x \alpha}
 H_{\alpha}^{-1}\, 
(H_{\alpha}^2+u^2)^{-1}
\, \ex^{-\im k_2\cdot x \alpha }H_{\alpha} (\vepsilon_2\cdot x)
\psi_{\alpha}
\big\ra\no[1ex]
&&\hspace{-10pt}\times 2 \big\la
\psi_{\alpha}, (\vepsilon_1\cdot x) H_{\alpha} \ex^{\im k_1\cdot x \alpha }
 H_{\alpha}^{-1}(H_{\alpha}^2+u^2)^{-1}\, 
\, \ex^{\im k_2\cdot x \alpha}H_{\alpha}  (\vepsilon_2\cdot x)
\psi_{\alpha}
\big\ra\,.
\end{eqnarray}
Next we use 
\begin{align}
H_{\alpha}(\vepsilon\cdot x)\psi_{\alpha}=[H_{\alpha}, \vepsilon\cdot
 x]\psi_{\alpha}=\im \vepsilon \cdot p \psi_{\alpha}
\end{align} 
and also introduce the integral kernel of
$(\one-P_\alpha)H_{\alpha}^{-1}(H_{\alpha}^2+u^2)^{-1}$
 as 
\begin{align}
K_{u, \alpha}(x, x')= \la x|(\one-P_\alpha)H_{\alpha}^{-1}(H_{\alpha}^2+u^2)^{-1}|x'\ra
\end{align} 
with $P_\alpha$ the projection onto $\psi_{\alpha}$. Note that $\la (\vepsilon\cdot p) \psi_{\alpha},  \ex^{\im k\cdot x}\psi_{\alpha}\ra=0$, which allows one to insert 
 $\one-P_\alpha$. 
Since $H_{\alpha}$ has a spectral gap, uniformly in $\alpha$, $K_{u,
\alpha}$
 is bounded  and $\la\phi, K_{u, \alpha} \phi\ra \cong C \la \phi,
 \phi\ra u^{-2}$ for $u\to \infty$.
With this notation we arrive at  the starting representation of $I_4$,
\begin{eqnarray}
&&\hspace{-10pt}I_4(\alpha^{-\gamma}R)
=\alpha^{6\gamma-6}(2\pi)^{-1} 4 \int_{\BbbR}\dm u \int_{\BbbR^6}\dm k_1\dm k_2\, 
|\hat{\varphi}(\alpha^{\gamma}k_1)|^2 |\hat{\varphi}(\alpha^{\gamma} k_2)|^2 \no[1ex]
&&\hspace{0pt}\times k_1^2(k_1^2+\alpha^{4-2\gamma}u^2)^{-1}
k_2^2(k_2^2+\alpha^{4-2\gamma}u^2)^{-1}\no
&&\hspace{0pt}\times 
\Big(\sum_{\lambda_1, \lambda_2}
\int_{\BbbR^{12}} \dm x \dm x' \dm y \dm y'
K_{u, \alpha}(x, x')K_{u, \alpha}(y,y')  \ex^{\im k_1\cdot (r+\alpha^{\gamma-1}(y-x))}
\ex^{\im k_2\cdot (r+\alpha^{\gamma-1}(y'-x'))}\no
&&\hspace{0pt} \times 
(\vepsilon_1\cdot p_x) (\vepsilon_1\cdot p_y)(\vepsilon_2\cdot p_{x'})
(\vepsilon_2\cdot p_{y'})
 \psi_{\alpha}(x) \psi_{\alpha}(y)
\psi_{\alpha}(x') \psi_{\alpha}(y')
\Big)\,.
\label{StartingI4}\medskip
\end{eqnarray} 
\textbf{Step 2} (Error estimate for vanishing phase).
We deal with the set on which the phase  in (\ref{StartingI4}) is close
to $0$ and  define 
\begin{align}
\Lambda_{r, \alpha}=\big\{(x,y)\in \BbbR^3\times \BbbR^3\, \Big|\, 
 |r+\alpha^{\gamma-1}(y-x)
|\ge \tfrac{1}{2}R
\big\}\,,
\end{align} 
correspondingly $\Lambda_{r, \alpha}' $ with $x,y$ replaced by $x', y'$.
$\tilde{I}_4$ is $I_4$ from (\ref{StartingI4}) with the integration
restricted to $\Lambda_{r, \alpha}\times \Lambda_{r, \alpha}'$.
The error term equals $I_4^{\mathrm{error}}=I_4-\tilde{I}_4$.
We use the Cauchy-Schwarz inequality inside (\ref{StartingI4}) and perform the 
$u, k_1, k_2$ integrations. Since $K_{u, \alpha}$ is bounded, this yields
\begin{align}
&\hspace{-25pt}|I^{\mathrm{error}}_4(\alpha^{-\gamma}R )| \le C \alpha^{-4\gamma-8} \int_{\BbbR^6} \dm k_1 \dm k_2 |k_1|^2 |k_2|^2 |\hat{\vphi}(k_1)|^2
|\hat{\vphi}(k_2)|^2\no
&\hspace{25pt} \times \Big(\int_{\BbbR^6\backslash \Lambda_{r, \alpha}} \dm
 x \dm y |\nabla \psi_{\alpha}(x)|^2|\nabla \psi_{\alpha}(y)|^2
\Big)^{1/2}\,.
\end{align} 
The ground state $\psi_{\alpha}$ has the exponential decay. Therefore 
\begin{align}
|I_4^{\mathrm{error}}(\alpha^{-\gamma}R)| \le C \alpha^{-4\gamma-8} \ex^{-\kappa R
 \alpha^{1-\gamma}}\,,
\end{align} 
which tends to $0$ as $\alpha\to 0$.
In the remainder we will study $\tilde{I}_4$.\medskip\\
\textbf{Step 3} (Angular integration).
The $k_1, k_2$ integrations are done in spherical coordinates setting
$\dm k_j=w_j^2 \dm w_j \dm \Omega_j,\ j=1,2$. 
Let $Q(k)=\one-|\hat{k}\ra\la \hat{k}|$ be the transverse projection.
 Then  the angular part reads 
\begin{align}
\int\dm \Omega_1 \, \ex^{\im k_1\cdot a_1} Q(k_1)\otimes 
\int\dm \Omega_2 \, \ex^{\im k_2\cdot a_2}Q(k_2)
\end{align} 
with
\begin{align}
a_1=r+\alpha^{\gamma-1}(y-x), \ \ \ a_2=r+\alpha^{\gamma-1}(y'-x').
\end{align} 
We omit the index and  compute 
$\displaystyle 
\int\dm \Omega\,  \ex^{\im k\cdot a } Q(k)
$ 
for general $a$.

Let $\mathsf{O}_a$ be an  orthogonal transformation in $\BbbR^3$ such that 
$\mathsf{O}_a a=|a|e_3$, where $e_3=(0,0,1)^T$.
Then 
\begin{align}
\int\dm \Omega\,  \ex^{\im k\cdot a }Q(k)&=\int\dm \Omega\,  \ex^{\im  k\cdot
 \mathsf{O}_a a }\mathsf{O}_a^{-1} Q(k) \mathsf{O}_a\no
&=2\pi\int_0^{\pi} \dm \vartheta \sin \vartheta 
\, \ex^{\im |k||a|\cos \vartheta }\mathsf{O}_a^{-1} \tilde{B}(\vartheta) \mathsf{O}_a\,,
\end{align} 
where $\tilde{B}_{ij}(\vartheta)=\delta_{ij} \tilde{b}_j(\vartheta),\
i,j=1,2,3$, with 
\begin{align}
\tilde{b}_1(\vartheta)=\tilde{b}_2(\vartheta)=\frac{1}{2}(1+(\cos\vartheta)^2),\
 \ 
\tilde{b}_3(\vartheta)=1-(\cos\vartheta)^2\,.
\end{align} 
Integrating over $\vartheta$ yields
\begin{align}
\int\dm \Omega\,  \ex^{\im k\cdot a }Q(k)=2\pi
 \mathsf{O}_a^{-1}B(|k||a|)\mathsf{O}_a\,, \label{AngInt}
\end{align} 
where $B_{ij}(s)=\delta_{ij}b_j(s)$ with 
\begin{align}
b_1(s)=b_2(s)=\hat{g}(s)-\hat{g}''(s)\,, \ \
 b_3(s)=2(\hat{g}(s)+\hat{g}''(s))\,,
\ \ \hat{g}(s)=s^{-1}\sin s\,.
\end{align} 
Thus, for $j=1,2$, 
\begin{align}
\int\dm \Omega_j\,  \ex^{\im k_j\cdot a_j }Q(k_j)=2\pi \mathsf{O}_{a_j}^{-1}B(|k_j||a_j|)\mathsf{O}_{a_j}\,.
\medskip
\end{align} 
\textbf{Step 4} (Radial integration).
The radial integrations are of the form
\begin{align}\label{pla}
\tfrac{1}{2}\int_{\BbbR} \dm w \hat{\varrho}(\alpha^{\gamma}w)w^4
 (w^2+\alpha^{4\gamma-2}u^2 )^{-1}f(|a||w|)
\end{align} 
with $f(s)=\hat{g}(s) = s^{-1}\sin s$ or $\displaystyle \hat{g}''(s)= 2 s^{-3}\sin s
- 2 s^{-2}\cos s - s^{-1}\sin s$. Here
$\hat{\varrho}(|k|)=|\hat{\varphi}(k)|^2$ and we extended $\hat{\varrho}$ to $\BbbR$ by reflection at $0$.
 We introduce a new function $\rho$ by
\begin{equation}
\rho(v) = (2\pi)^{-1/2}\int_{\BbbR} \dm w  \hat{\varrho}(w) \ex^{\im  v w}.
\end{equation}
 Then one has
\begin{equation}
\int_{\BbbR}\dm v\rho(v) = (2\pi)^{1/2}\hat{\varrho}(0) =   (2\pi)^{-5/2}\,.
\end{equation}
Let $\sigma(|x|) = \varphi \ast \varphi(x)$. Then
\begin{equation}
\rho(v) = (2\pi)^{-3/2}\int_v^\infty \dm r\,r\sigma(r)
\end{equation}
for $v \geq 0$. Since $\varphi$ is continuous and of compact support, $\rho \in C^1( \BbbR)$ and $\rho$ has
has compact support.

We plan to use Plancherel's theorem in (\ref{pla}) and obtain, in the sense of distributions,
\begin{align}
c_1(v; |a|, u)&=(2\pi)^{-1/2}\int_{\BbbR}\dm w\,  \ex^{\im  v w}w^3
 (w^2+u^2)^{-1}|a|^{-1} \sin (|a|w)\no
&=(4|a|)^{-1}(2\pi)^{1/2}\Big(
-2\delta'(v+|a|)-u^2 \mathrm{sgn}(v+|a|)\, \ex^{-|u||v+|a||}\no
&\hspace{20pt}+2\delta'(v-|a|)+u^2 \mathrm{sgn}(v-|a|)\, \ex^{-|u||v-|a||}
\Big)\,,\label{c1}\\
c_2(v; |a|, u)&=(2\pi)^{-1/2}\int_{\BbbR}\dm w\,  \ex^{\im  v w}w
 (w^2+u^2)^{-1} |a|^{-3}\sin (|a|w)\no
&=(4|a|^3)^{-1}(2\pi)^{1/2}\Big(
\mathrm{sgn}(v+|a|)\, \ex^{-|u||v+|a||}- \mathrm{sgn}(v-|a|)\, \ex^{-|u||v-|a||}
\Big)\,,\label{c2}\\
c_3(v; |a|, u)&=(2\pi)^{-1/2}\int_{\BbbR}\dm w\,  \ex^{\im v w}w^2
 (w^2+u^2)^{-1}|a|^{-2} \cos (|a|w)\no
&=(4|a|^2)^{-1}(2\pi)^{1/2}\Big(
2\delta(v+|a|)-|u| \, \ex^{-|u||v+|a||}\no
&\hspace{20pt}+2\delta(v-|a|)-|u| \, \ex^{-|u||v-|a||}
\Big)\label{c3}
\end{align}  
with the sign function $\mathrm{sgn}(t)=1$ for $t \ge 0$, $\mathrm{sgn}(t)=-1$ for $t<0$.
Since $\rho \in C^1$, Plancherel's theorem yields
\begin{align}
d_1(|a|, u, \alpha)&= \int_{\BbbR}\dm v \alpha^{-\gamma}
 \rho(\alpha^{-\gamma}v ) \big(
c_1(v;|a|, u)-c_2(v; |a|, u)+c_3(v; |a|, u)
\big),\nonumber\\[1ex]
d_2(|a|, u, \alpha)&=d_1(|a|, u, \alpha)\,,\nonumber\\[1ex]
d_3(|a|, u, \alpha)&= \int_{\BbbR}\dm v \alpha^{-\gamma}
 \rho(\alpha^{-\gamma }v)2 \big(
c_2(v; |a|, u)-c_3(v;|a|, u)
\big)
\end{align} 
and 
\begin{align}
D_{ij}(|a|, u, \alpha)=(2\pi)\delta_{ij}d_j(|a|, u, \alpha)\,.
\end{align} 

We now combine all terms. The ground state $\psi_{\alpha}$ is invariant
under rotations. It is convenient to write 
\begin{align}
\psi_{\alpha}(x)=\psi_{\alpha, \mathrm{rad}}(|x|)\,, \ \ \
 \nabla\psi_{\alpha}(x)=\psi'_{\alpha, \mathrm{rad}}(|x|)\hat{x}\,,
\end{align} 
where $\hat{x}=x/|x|$. Then (\ref{StartingI4}) becomes
\begin{eqnarray}
&&\hspace{-30pt}\tilde{I}_4(\alpha^{-\gamma}R)\no
&&\hspace{-15pt}=\alpha^{6\gamma-6}(2\pi)^{-1} 4 \int_{\BbbR}\dm u 
\int_{\Lambda_{r, \alpha}\times \Lambda'_{r, \alpha}} \dm x \dm y \dm x'
\dm y'
\no
&&\hspace{0pt}\times
\Big(
\hat{x}\cdot  \mathsf{O}^{-1}_{r+\alpha^{\gamma-1}(y-x)} D\big(
 |r+\alpha^{\gamma-1}(y-x)|, \alpha^{2-\gamma}u, \alpha\big)
\mathsf{O}_{r+\alpha^{\gamma-1}(y-x)} \hat{y}
\Big)
\no[1ex]
&&\hspace{0pt}\times
\Big(
\hat{x}'\cdot  \mathsf{O}^{-1}_{r+\alpha^{\gamma-1}(y'-x')}
 D\big(|r+\alpha^{\gamma-1}(y'-x')|, \alpha^{2-\gamma}u, \alpha\big)
\mathsf{O}_{r+\alpha^{\gamma-1}(y'-x')}\hat{y}'
\Big)\no[1ex]
&&\hspace{0pt}\times {\psi}'_{\alpha,
 \mathrm{rad}}(|x|){\psi}'_{\alpha,
 \mathrm{rad}}(|x'|){\psi}'_{\alpha,
 \mathrm{rad}}(|y|){\psi}'_{\alpha, \mathrm{rad}}(|y'|)
K_{u, \alpha}(x, x')K_{u, \alpha}(y, y')\,.
\label{ResultingI4}\medskip
\end{eqnarray} 
\textbf{Step 5} (The limit $\alpha\to 0$). $c_1, c_2, c_3$ contain terms proportional to
$\delta$ and $\delta'$.  
Since, by assumption, $\varrho$ has compact support and since $|a_j|$ is bounded away from zero on the 
prescribed domain of integration, these terms vanish for $\alpha$ sufficiently small. 
Thus only the regular terms, containing
the exponential function, have still to be considered.

We have to discuss the cases $1<\gamma <2$ and $\gamma\ge 2$ separately.
\medskip\\
$1<\gamma <2$. As before we use the uniform bound from
$K_{u, \alpha}$. Therefore the terms proportional  to $u^0, u, u^2$
 have a uniformly integrable bound  in $u$. By dominated convergence 
only the term proportional  to $u^0$ does not vanish as $\alpha\to 0$.
The term proportional to $u^3, u^4$ are bounded  as 
\begin{align}
(1+u^2)^{-2} \big(
(\alpha^{2-\gamma} u)^4+(\alpha^{2-\gamma }|u|)^3
\big)\, \ex^{-\kappa \alpha^{2-\gamma}|u|}
\le C \alpha^{3(2-\gamma)}\, \ex^{-\kappa (2-\gamma)|u|}
\end{align} 
with $\kappa\ge \kappa_0>0$ uniformly in $\alpha$. 
Thus the integral over $u$ vanishes as $\alpha\to 0$.

We are left with the products of the $u^0$ terms. As $\alpha\to 0$, the
matrix $\mathsf{O}_{r+\alpha^{\gamma-1}(y-x)}$ tends to the unit
matrix. 
Thus we conclude 
\begin{align}
&\hspace{-10pt}\lim_{\alpha\to 0} \alpha^{6-6\gamma}\tilde{I}_4(\alpha^{-\gamma} R)\no
&=(2\pi)^{-1} 4 \Big(
(2\pi)^{3/2} (2R^3)^{-1}\tfrac{1}{2}\int_{\BbbR} \dm v \rho(v) 
\Big)^2\no
&\hspace{20pt}\times\int_{\BbbR}\dm u 
\int_{\BbbR^{12}}
\dm x\dm x'\dm y\dm y'
\big(
\hat{x}_1\hat{y}_1+\hat{x}_2\hat{y}_2+2\hat{x}_3\hat{y}_3
\big)
\big(
\hat{x}'_1\hat{y}'_1+\hat{x}'_2\hat{y}'_2+2\hat{x}'_3\hat{y}'_3
\big)\no[1ex]
&\hspace{20pt}\times {\psi}'_{\mathrm{hy},
 \mathrm{rad}}(|x|){\psi}'_{\mathrm{hy},
 \mathrm{rad}}(|x'|){\psi}'_{\mathrm{hy},
 \mathrm{rad}}(|y|){\psi}'_{\mathrm{hy}, \mathrm{rad}}(|y'|)
K_{u, 0}(x, x')K_{u, 0}(y, y')\,. \label{1gamma2}
\end{align} 
Using the rotational invariance of $K_{u, 0}$ one arrives at 
\begin{align}
&\hspace{-5pt}\lim_{\alpha\to 0} \alpha^{6-6\gamma} \tilde{I}_4(\alpha^{-\gamma}R)
= (2\pi)^{-3} 2^{-1}3 R^{-6} \int_{\BbbR} \dm u \big(
\tfrac{1}{3}\big\la\psi_{\mathrm{hy}}, x\cdot
 H_{\mathrm{hy}}(H_{\mathrm{hy}}^2
+u^2)^{-1} x \psi_{\mathrm{hy}}\big\ra
\big)^2\no
&\hspace{40pt}= (2\pi)^{-2}2^{-2} a_{\mathrm{VW}}R^{-6}.\bigskip
\end{align} 
$2\le \gamma$. We substitute $u$ by
$\alpha^{\gamma-2}u$. 
 The uniform bound now  results from the
exponential terms $\exp[-|u|v\pm |a|]$, using that  
\begin{align}
\int_{\BbbR}\dm v |\rho(v)|\ex^{-|u||\alpha^{\gamma}v\pm |a||} \le C
 \ex^{-\kappa |u|}
\end{align} 
uniformly in $\alpha$, provided $\alpha$ is sufficiently small, since
 $\rho(v)$  has compact support by the assumption.  In the limit
$\alpha\to 0$ one obtains  a formula  which has the same structure as in
(\ref{1gamma2}). Only the coefficients in front of $\hat{x}_j,
\hat{y}_j$ and  $\hat{x}_j',
\hat{y}_j'$ 
 are now different.

For $\gamma=2$, one obtains 
\begin{align}
&\lim_{\alpha\to 0} \alpha^{-6} \tilde{I}_4(\alpha^{-2}R)=(2\pi)^2
\Big(
\int_{\BbbR}\dm v \rho(v)
\Big)^2 \no
&\hspace{50pt}\times \int_{\BbbR}\dm u \Big(
\beta_1^2+3\beta_2^2+3\beta_3^2-2\beta_1 \beta_2-6 \beta_2
 \beta_3+2\beta_1 \beta_3
\Big)
2\ex^{-2 R|u|}\no
&\hspace{50pt}\times \Big(
\tfrac{1}{3}\big\la
\psi_{\mathrm{hy}}, x\cdot (\one-P_{\mathrm{hy}})H_{\mathrm{hy}}(H_{\mathrm{hy}}^2+u^2)^{-1}x
\psi_{\mathrm{hy}}
\big\ra
\Big)^2
\end{align}
with $\beta_1=-R^{-1} u^2,\ \ \beta_2=R^{-3},\ \ \beta_3=-R^{-2}|u|$,
which yields 
\begin{align}
\lim_{\alpha\to 0 }\alpha^{-6} \tilde{I}_4(\alpha^{-2}R)=(2\pi)^{-2
 }2^{-1} h_{\mathrm{co}}(R)\,.
\end{align} 

For $\gamma>2$, one has the same expression except that from the rescaling of $\dm u$ one picks up the factor $\alpha^{\gamma-2}$ and that the factor 
$(\one-P_{\mathrm{hy}})(H_{\mathrm{hy}}^2+u^2)^{-1}$ now reads $(\one-P_{\mathrm{hy}})(H_{\mathrm{hy}}^2
+(\alpha^{\gamma-2 }u)^2)^{-1}$, which is still uniformly bounded in $u$. Hence 
\begin{align}
\lim_{\alpha\to 0} \alpha^{8-7\gamma}\tilde{I}_4(\alpha^{-\gamma}R)
=(2\pi)^{-2}2^{-1} a_{\mathrm{CP}} R^{-7}\,.
\end{align}  
This concludes the proof of Proposition \ref{AssertionI4}. $\Box$

%%%%%%%%%%%%%%%%%%%%%%%%%%%%%%%%%%%%%%%%%%
%%%%%%%%%%%%%%%%%%%%%%%%%%%%%%%%%%%%%%%%%%%%

\section{Strong resolvent convergence}
\setcounter{equation}{0}

We discuss the limit $\alpha\to 0$ for a free electron coupled to the
radiation field on the scale set by the hydrogen atom.
Then the energies are of order $\alpha^2$ and hamiltonian on that scale
reads
\begin{align}
T_{1,\alpha}=\tfrac{1}{2}:\hspace*{-3pt} \big(p-\sqrt{4\pi}
 \alpha^{3/2}A_{\alpha}(x)\big)^2\hspace*{-3pt}: +\Hf\,. \label{SingleEL}
\end{align} 
The coupling function in (\ref{SingleEL}) is 
\begin{align}
g_{\alpha}(k, \lambda)=\sqrt{4\pi}\alpha^{3/2}\hat{\vphi}(\alpha^2
 k)\frac{1}{\sqrt{2\omega }}\ex^{\im \alpha k\cdot x} \vepsilon(k, \lambda).
\end{align} 
Note that $\|g_{\alpha}\|\cong \alpha^{-1/2} $, which makes the limit
$\alpha\to 0$ singular. 

The ultraviolet cutoff as $\alpha^{-2}$ corresponds to a charge distribution localized on the relativistic scale.
If the charge distribution would have a width of the order of the Bohr radius, then 
$\hat{\vphi}(\alpha^2 k)$ would have to be replaced by $\hat{\vphi}(\alpha k)$. Thus it is natural to introduce 
the parameter $\delta$ with $0 \leq \delta < 1$ and to define $T_{1,\alpha}^{(\delta)}$ by (\ref{SingleEL}) with
$\hat{\vphi}(\alpha^2 k)$ substituted through $\hat{\vphi}(\alpha^{2 - \delta} k)$. If $0 < \delta < 1$, our arguments in Sections 2 and 3 would not be altered, except for $a_0 = 0$. But now the resolvent convergence can be established.
\begin{Prop}\label{resolvent} Let $0 < \delta < 2$. Then, in the sense of strong convergence of resolvents,
\begin{equation}
\lim_{\alpha\to 0} T_{1,\alpha}^{(\delta)} =\tfrac{1}{2}p^2 +\Hf\,. 
\end{equation} 
\end{Prop}
\textbf{Proof}: Let
\begin{align}
T_{1. \alpha}^{(\delta)}=\tfrac{1}{2}:\hspace*{-3pt}\big(p-\sqrt{4\pi}\alpha^{3/2}A_{\alpha, \delta}(x)\big)^2\hspace*{-3pt}:+\Hf
=T_0+B_{\alpha, \delta}\,,
\end{align} 
where
\begin{eqnarray}
&&\hspace{-38pt}T_0=\tfrac{1}{2}p^2+\Hf\,,\no
&&\hspace{-38pt} B_{\alpha, \delta}=-\sqrt{4\pi}\alpha^{3/2}p\cdot (A^+_{\alpha, \delta}(x)+A_{\alpha,\delta}^-(x))\no
&&\hspace{0pt}+2\pi\alpha^3 \big(
A_{\alpha,\delta}^+(x)\cdot A_{\alpha,\delta}^+(x)+
2A_{\alpha,\delta}^+(x)\cdot A_{\alpha,\delta}^-(x)+
A_{\alpha,\delta}^-(x)\cdot A_{\alpha,\delta}^-(x)
\big)\,.
\end{eqnarray} 
The coupling function of $A_{\alpha, \delta}(x)$ is given by
\begin{align}
g_{\alpha, \delta}(k, \lambda)=\hat{\vphi}(\alpha^{2-\delta}
 k) \frac{1}{\sqrt{2\omega}} \ex^{\im \alpha k\cdot x }\varepsilon(k, \lambda)\,.
\end{align} 

If it can be shown that
\begin{equation}\label{Bound}
|\la \phi, B_{\alpha, \delta} \phi\ra|\le
  C(\alpha)\|(T_0+\one)^{1/2}\phi\|\,, 
\quad C(\alpha)\to 0 \ \ \mathrm{as}\ \  \alpha\to 0\,,
\end{equation} 
for all $\phi\in \D(T_0^{1/2})$,
then one concludes 
$
T_{1, \alpha}^{(\delta)}\to \tfrac{1}{2}p^2+\Hf$ as $\alpha\to 0
$
in the norm resolvent sense by the general theorem \cite[Theorem VIII.25]{ReSi1},
as based on  the famous Nelson's argument \cite{Nelson}.
To prove (\ref{Bound}), we apply the standard bounds
\begin{align}
\|a(f) \psi\|&\le \|\omega^{-1/2}f\|\|(\Hf+\one)^{1/2}\psi\|\,,
 \label{Anni}\\[1ex]
\|a(f)^*\psi\|^2& \le (\|f\|^2+\|\omega^{-1/2} f\|^2) \|(\Hf+\one)^{1/2}
 \psi\|^2\,. \label{Crea}
\end{align} 
As to $A_{\alpha, \delta}^-(x)$, the bound (\ref{Anni}) translates to
\begin{align}
\alpha^{3/2} \|A^-_{\alpha, \delta}(x)\psi\|&\le \alpha^{3/2}
 \|\omega^{-1/2} g_{\alpha, \delta}\|\|(T_0+\one)^{1/2}\psi\|\\
& \le  \mathcal{O}(\alpha^{(1+\delta)/2}) \|(T_0+\one)^{1/2}\psi\|\,.
\end{align} 
Similarly  $A^-_{\alpha, \delta}\cdot A^-_{\alpha, \delta}$
can be estimated as 
\begin{eqnarray}
&&\hspace{-30pt}\alpha^3\|(\Hf+\one)^{-1/2}A^-_{\alpha, \delta}(x)\cdot A^-_{\alpha,
 \delta}(x)\psi\|\no
&&\hspace{0pt}\le
\alpha^3 \|A_{\alpha, \delta}^+ (\Hf+\one)^{-1/2}\| \|A_{\alpha,
 \delta}^-(\Hf+\one)^{-1/2}\| \|(\Hf+\one)^{1/2}\psi\|\no
&&\hspace{0pt}\le \alpha^3 (\|g_{\alpha, \delta}\|^2+\|\omega^{-1/2} g_{\alpha,
 \delta}\|^2)^{1/2} \|\omega^{-1/2} g_{\alpha, \delta}\|
 \|(\Hf+\one)^{1/2}\psi\|\no
&&\hspace{0pt}\le \mathcal{O}(\alpha^{3\delta/2}) \|(\Hf+\one)^{1/2}\psi\|\,.
\end{eqnarray} 
Thus we arrive at 
\begin{align}
\alpha^{3/2}| \la \phi, p\cdot A_{\alpha, \delta}^- \phi\ra|
\le \mathcal{O}(\alpha^{(1+\delta)/2}) \|(T_0+\one)^{1/2}\phi\|^2
\end{align} 
and
\begin{align}
\alpha^3 |\la \phi, A_{\alpha, \delta}^-(x)\cdot A_{\alpha, \delta}^-(x)
 \phi\ra|
&\le \alpha^3 \|(\Hf+\one)^{1/2}\phi\|\|(\Hf+\one)^{-1/2} A^-_{\alpha,
 \delta}\cdot A^-_{\alpha, \delta}
 \phi\|\no
&\le \mathcal{O}(\alpha^{3\delta/2}) \|(T_0+\one)^{1/2}\phi\|^2
\end{align} 
for each $\phi\in \D(T_0^{1/2})$.
Hence (\ref{Bound}) is satisfied and the assertion follows.  \medskip$\Box$

The case $\delta =0$ is physically distinguished, but Nelson's argument fails and no other 
functional analytic method seems to be available.  
We devise an alternative approach based on functional integrals, which clearly displays that
$\delta = 0$ is on the borderline.  In our context functional integration is explained in 
\cite{Spohn}, Chapter 14, and at greater depth in \cite{BHL}. The
propagator $\ex^{-\tau T_{1,\alpha}}$ can be written as an integral with
respect to Brownian motion for the particle and a Gaussian space-time
measure  for the Maxwell field. It is convenient to pick for $\psi$
the particular form 
\begin{align}
\psi=\phi\otimes W(f)\Omega\,.
\end{align} 
Here $\phi\in L^2(\BbbR^3)$, $\Omega$ is the Fock vacuum, and $W(f)$ is
the Weyl operator 
\begin{align}
W(f)=\ex^{(a(f)^*-a(f))}\,,\ \ f\in L^2(\BbbR^3)\otimes \BbbC^2\,.
\end{align} 
Note that the linear span of these vectors is dense in
$L^2(\BbbR^3)\otimes \Fock$.
Integrating over the Maxwell field one arrives at the following
expression
\begin{align}
\la \psi, \ex^{-\tau T_{1,\alpha}}\psi\ra=\mathbb{E}_{\mathrm{W}}
\big(
\phi(q_0)^* \phi(q_{\tau})\ \ex^{-\mathscr{A}}
\big)\,,
\end{align} 
where $t\mapsto q_t$ is a path in $\BbbR^3$ and
$\mathbb{E}_{\mathrm{W}}$ denotes average over the Wiener measure.
The action $\mathscr{A}$ results from the Gaussian integration over the
Maxwell field and consists of a sum of three pieces,
\begin{align}
\mathscr{A}=\mathscr{A}_1+\mathscr{A}_2+\mathscr{A}_3\,.
\end{align}

$\mathscr{A}_1$ is the piece corresponding to $f=0$,
\begin{align}
\mathscr{A}_1=4\pi\alpha^3 \int_0^{\tau}\int_0^t\dm q_t\cdot
 W_{\alpha}(q_t-q_s, t-s)\, \dm q_s \label{Action1}
\end{align} 
with the photon propagator
\begin{align}
W_{\alpha}(x,t)=\int_{\BbbR^3}\dm k |\hat{\vphi}(\alpha^2 k)|^2
 \frac{1}{2\omega(k)}
\ex^{\im k\cdot x\alpha} \ex^{-\omega(k)|t|} Q(k)
\end{align}
and $Q(k)=\one -|\hat{k}\ra\la\hat{k}|$, the transverse projection.
(\ref{Action1}) is an iterated Ito integral. It avoids the diagonal
$\{s=t\}$ in accordance with the Wick ordering $:\  :$.
$\mathscr{A}_3$ reflects the term coming from $W(f)$,
\begin{align}
\mathscr{A}_3=\int_{\BbbR^3}\dm k\frac{1}{\omega}(1+\ex^{-\omega \tau})
\hat{f}^*(k)\cdot Q(k)\hat{f}(k)\,.
\end{align} 
Note that $\mathscr{A}_3$ does not depends on $q_t$ and $\alpha$. Finally
the cross term $\mathscr{A}_2$ reads 
\begin{align}
\mathscr{A}_2=-\im \sqrt{4\pi}\alpha^{3/2}\int_0^{\tau}\int_{\BbbR^3}\dm k\, 
 \hat{\vphi}(\alpha^2 k) \frac{1}{\sqrt{2\omega}} \ex^{\im k\cdot q_t
 \alpha}
\big(
\ex^{-\omega t}+\ex^{-\omega(\tau-t)}
\big)\, \dm q_t\cdot Q(k)\hat{f}(k)\,. 
\end{align} 

The cross term is small, since for the expectation $\mathbb{E}_0$  with respect to standard
Brownian motion starting at $q_0=0$ it holds 
\begin{eqnarray}
&&\hspace{-40pt}\mathbb{E}_0\big( |\mathscr{A}_2|^2\big) =4\pi \alpha^3 \int_0^{\tau}\int_{\BbbR^6}\dm
 k_1 \dm k_2\, \hat{\vphi}(\alpha^2 k_1)\hat{\vphi}(\alpha^2 k_2)
 (2|k_1|2|k_2|)^{-1/2}\no
&&\hspace{50pt}\times \ex^{-(\alpha^2 \frac{1}{2}(k_1+k_2)^2+|k_1|+|k_2|)t}
\hat{f}(k_1)\cdot Q(k_1)Q(k_2)\hat{f}(k_2)\no
&&\hspace{12pt}\le \alpha \pi \|\hat{\vphi}\omega^{-1}\|^2 \|\hat{f}\|^2\,.
\end{eqnarray} 
Ignoring the cross term one arrives at 
\begin{align}
\big\la \phi\otimes W(f)\Omega, \ex^{-\tau T_{1,\alpha}}\phi\otimes
 W(f)\Omega\big\ra
=\mathbb{E}_{\mathrm{W}}\Big(
\phi(q_0)^* \phi(q_{\tau}) \, \ex^{-\mathscr{A}_1}
\Big)\, \ex^{-\mathscr{A}_3}+\mathcal{O}(\alpha)\,. \label{Expectation}
\end{align} 
It is now convenient to rewrite (\ref{Expectation}) using that
$\mathscr{A}_1$ depends only on the increments. Then
\begin{align}
\big\la \phi\otimes W(f)\Omega, \ex^{-\tau T_{1,\alpha}}\phi\otimes
 W(f)\Omega\big\ra&=\int_{\BbbR^3}\dm k |\hat{\phi}(k)|^2
 \mathbb{E}_0\big(
\ex^{-\mathscr{A}_1}\ex^{\im k\cdot q_{\tau}}
\big) \, \ex^{-\mathscr{A}_3}+\mathcal{O}(\alpha)\,,
\end{align}

To turn to $\mathscr{A}_1$ we first note that
$\mathbb{E}_0(\mathscr{A}_1)=0$ by Ito calculus.
Secondly we use the Brownian motion scaling $q_t=\alpha q_{t/\alpha^2}$
 to rewrite $\mathscr{A}_1$ as 
\begin{align}
\mathscr{A}_1=4\pi\alpha \int_{0\le s<t\le \tau/\alpha^2} \dm q_t\cdot
 W_1(q_t-q_s, t-s)\, \dm q_s\,.
\end{align} 
By definition $W_1$ does not depend on $\alpha$. $W_1$ decays as
$(t-s)^{-2}$, which should provide enough independence  for a central
limit theorem hold.
Thus the key input is that $(\mathscr{A}_1, q_{\tau})$ jointly converge
to a Gaussian as $\alpha\to 0$. One checks that 
\begin{align}
\mathbb{E}_0(\mathscr{A}_1q_{\tau})=0\,.
\end{align} 
Hence the assumption is that 
\begin{align}
\lim_{\alpha\to 0}\mathscr{A}_1=\xi_{\mathrm{G}}
\end{align} 
with $\xi_{\mathrm{G}}$ a centered Gaussian random variable independent
of $q_{\tau}$. To complete our argument we compute the variance of
$\mathscr{A}_1$,
\begin{eqnarray}
&&\hspace{-20pt}\mathbb{E}_0\big(\mathscr{A}_1^2\big)\no
&&\hspace{0pt}=(4\pi \alpha)^2 \int_{0\le s_1<t_1\le
 \tau/\alpha^2}
\int_{0\le s_2<t_2\le \tau/\alpha^2}\mathbb{E}_0\big(
\big(\dm q_{t_1}\cdot W_1(q_{t_1}-q_{s_1}, t_1-s_1)\, \dm
 q_{s_1}\big)\no
&&\hspace{186pt}\times \big(
\dm q_{t_2}\cdot W_1(q_{t_2}-q_{s_2}, t_2-s_2)
\, \dm q_{s_2}\big)
\big)\no
&&\hspace{0pt}=(4\pi\alpha)^2\int_{0\le s<t\le \tau/\alpha^2} \mathbb{E}_0
\big(
\Tr\big(
W_1(q_t-q_s, t-s)^2
\big)
\big)\, \dm s\dm t\,.
\end{eqnarray} 
Hence 
\begin{eqnarray}
&&\hspace{-50pt}\lim_{\alpha\to 0}\mathbb{E}_0\big(\mathscr{A}_1^2)=(4\pi\big)^2
 \int_0^{\infty}
\dm t \int_{\BbbR^6}\dm k_1\dm k_2\, |\hat{\vphi}(k_1)|^2|\hat{\vphi}(k_2)|^2
(2|k_1|2|k_2|)^{-1}\no
&&\hspace{26pt}\times \exp\Big\{
-\frac{1}{2}\Big(
(k_1+k_2)^2+|k_1|+|k_2|
\Big)t
\Big\}\Tr(Q(k_1) Q(k_2))\no
&&\hspace{12pt}=2a_0 \tau\,.
\end{eqnarray} 
Returning to (\ref{Expectation}) one concludes that 
\begin{eqnarray}
&&\hspace{-40pt}\lim_{\alpha\to  0}\big\la
\phi\otimes W(f)\Omega, \ex^{-\tau T_{1,\alpha}}\phi\otimes W(f)\Omega
\big\ra
=\int_{\BbbR^3}\dm k\, |\hat{\phi}(k)|^2 \mathbb{E}_0(\ex^{\im k\cdot
 q_{\tau}}) \mathbb{E}(\ex^{\xi_{\mathrm{G}}})\, \ex^{-\mathscr{A}_3}\no
&&\hspace{40pt}=\big\la
\phi\otimes W(f)\Omega, \ex^{-\tau ((p^2/2)+\Hf-a_0)}
\phi\otimes W(f)\Omega
\big\ra\,.
\end{eqnarray} 

If one reintroduces the parameter $\delta$ from above, then the variance vanishes provide $\delta > 0$, in accordance with Proposition 4.1.

\end{document}